
\input harvmac

\def\nl{\hfil\break}

\def\npb{{ \sl Nucl. Phys. }}

\def\prd{{ \sl Phys. Rev. }}
\def\prl{{ \sl Phys. Rev. Lett. }}
\def\plb{{ \sl Phys. Lett. }}

\def\undertext#1{\vtop{\hbox{#1}\kern 1pt \hrule}}
\def\half{{1\over2}}
\def\c#1{{\cal{#1}}}
\def\dirac{\hbox{$\partial$\kern-0.5em\raise0.3ex\hbox{/}}}
\def\dslash{\hbox{$\partial$\kern-0.5em\raise0.3ex\hbox{/}}}
\def\Dirac{\hbox{{\mit D}\kern-0.6em\raise0.3ex\hbox{/}}}
\def\Dslash{\hbox{{\mit D}\kern-0.6em\raise0.3ex\hbox{/}}}
\def\kslash{\hbox{{\mit k}\kern-0.4em\raise0.3ex\hbox{/}}}
\def\pslash{\hbox{{\mit p}\kern-0.5em\raise-0.3ex\hbox{/}}}
\overfullrule=0pt
\def\nc{N_{c}}
\def\sgn{\,{\rm sign}\,}
\def\ph#1{\phi_{#1}}
\def\ps#1{\psi_{#1}}
\def\lm{\lambda_-}
\def\nf{N_{flavors}}
\nopagenumbers\abstractfont\hsize=\hstitle\rightline{TIT/HEP--230}
\vskip 1in\centerline{\titlefont
Boson--fermion bound states in two dimensional QCD
}
\abstractfont\vskip .5in\pageno=0
\centerline{Kenichiro Aoki\footnote{$^\dagger$}{%
email:{\tt~ken@phys.titech.ac.jp}}}
\bigskip\centerline{\it Department of Physics}
\centerline{\it Tokyo Institute of Technology}
\centerline{\it Oh-Okayama, Meguro-ku}
\centerline{\it Tokyo, JAPAN  152}
\vskip .3in
\baselineskip=12pt plus 2pt minus 1pt
\centerline{\bf Abstract}
We derive the boson--fermion bound state equation
in a two dimensional gauge theory in the large--$\nc$ limit.
We analyze the properties of this equation and
in particular, find that the mass trajectory is
linear with respect to the bound state level
for the higher mass states.

\Date{7/93}
\newsec{Introduction}
It is quite common in particle physics to encounter the
possibility of boson--fermion bound states.
This possibility often arises in technicolor theories,
especially when we have supersymmetry.
Also, it should not be forgotten that in order
to understand the dynamics of the standard electroweak
model, we need to consider also the region where the gauge
couplings are strong and quarks and leptons are thought to be
as composite particles \ref\AF{G.~{}'t Hooft,
Carg\`ese Summer Inst. Lectures, Plenum Press, NY (1980);\nl
L.~Abbott, E.~Farhi, \plb{\bf 101B} (1981) 69,
\npb{\bf B189} (1981) 547}%
{}.
Compositeness, in essence, is a consequence of non--perturbative
phenomena and even in the most studied case, QCD, it is still considered
an outstanding problem.
In this regard, the large--$\nc$ limit of QCD in two dimensions is
to a large extent solvable \ref\THOOFT{G.~{}'t Hooft,
\npb{\bf B72} (1974) 461, {\bf B75} (1974) 461} and presents an invaluable
opportunity for studying QCD in a simpler situation
\ref\TWODQCD{C.G.~Callan, N.~Coote, D.J.~Gross, \prd{\bf D13}
(1976) 1649;\nl
M.B.~Einhorn, \prd{\bf D14} (1976) 3451, {\bf D15} (1976) 3037;\nl
R.C. Brower, J.~Ellis, M.G.~Schmidt, J.H.~Weiss, \plb{\bf 65B} (1976) 249
}%
\ref\SHEI{S.S.~Shei, H.S. Tsao, \npb{\bf B141} (1978) 445}.
Substantial amount of work has been done using this model which has
yielded important physical insight into the dynamics of QCD.

In this note, we derive the boson--fermion bound state equation
in the large--$\nc$ limit of two dimensional QCD
analogous to the fermion--fermion \THOOFT\ and the boson--boson
bound state equations \SHEI\ref\BP{W.A.~Bardeen,
P.B.~Pearson, \prd{\bf D14} (1976) 547;\nl
M.B.~Halpern, P.~Senjanovic, \prd{\bf D15} (1977) 1655}
derived previously.
The properties of this equation will be examined
and it is shown that the higher mass bound states
follow a trajectory linear with respect to the bound state level.

\def\mb#1{m_{(b)#1}}
\def\mf#1{m_{(f)#1}}
The Lagrangian of the model is
\eqn\lag{-\c L={1\over4}\tr(F_{\mu\nu}^2)
+\sum_a\overline\psi_a\left(\Dslash+\mf a\right)\psi_a
+\sum_a\left(\left|D\phi_a\right|^2 +\mb a^2\left|\phi_a\right|^2\right)
}
$a$ denotes the flavor index and $(f),(b)$ indices are used
to indicate quantities pertaining respectively to
fermions and bosons.
The space--like metric $(-+)$ will be used in this work.
We choose the gauge group to be SU$(\nc)$ and the matter
fields $\ph a,\ps a$ to belong to the fundamental representation.
We adopt the light--cone gauge $A_-=A^+=0$.
The light--cone components are defined as $a^\pm=a_\mp\equiv
 1/\sqrt2(a^1\pm a^0)$.

The large--$\nc$ limit is taken by fixing $g^2\nc$ and the masses
while taking $\nc$ to infinity.
The leading order quantum corrections to the
propagators are of $\c O(\nc^0)$ and
we may solve the Schwinger--Dyson equations for the
fermion and the boson
propagators incorporating these contributions as,
respectively,
\eqn\props{\eqalign{
S_a(p) &= \left[i\pslash+i{g^2\nc\over2\pi}\left({\sgn(p_-)\over\lm}
-{1\over p_-}\right)\gamma^++\mf a\right]^{-1}\cr
D_a(p) &=\left[p^2+\mb a^2+{g^2\nc\over\pi}{|p_-|\over\lm}\right]^{-1}\cr
}}
Here, $\lm$  denotes the infrared cutoff.
The Schwinger--Dyson equations may be represented graphically as in
\fig\figprop{}.\nl
\figprop{\ \it Schwinger--Dyson
equations for the self--energy part of the propagators.}\medskip\noindent
\def\mubf{\mu_{(bf)}}
\def\muff{\mu_{(ff)}}
\def\mubb{\mu_{(bb)}}
\def\alf#1{\alpha_{(f)#1}}
\def\alb#1{\alpha_{(b)#1}}
\def\tpbf{\tilde\varphi_{(bf)}}
\def\tpbb{\tilde\varphi_{(bb)}}
\def\tpff{\tilde\varphi_{(ff)}}

\def\pbf{\varphi_{(bf)}}

\def\psbf{\psi_{(bf)}}
\def\pint{{\rm P}\!\!\!\int_0^1\!\!\!{dy\over(y-x)^2\,}}
The Bethe--Salpeter equation for the bound states
to leading order in $1/\nc$
may be represented diagrammatically as \fig\figbs{}
and may be obtained following {}'t Hooft \THOOFT;
\eqn\bsorig{\psbf(p,r)=g^2D_1(p)S_2(p-r)\gamma^+
\int\!{d^2\!k\over(2\pi)^2}\,{(2p_-+k_-)\over k_-^2}
\psbf(k+p,r)}
\figbs{\ \it Bethe--Salpeter equations for bound states.
The blobs denote the connected parts of the diagram.}\medskip\noindent
In the light--cone gauge, using $(\gamma^+)^2=0$,
$\psbf(p,r)$ is proportional to $\gamma^-$
so that we define
$\int\!dp_+\,\psbf(p,r)\equiv\gamma^-\tpbf(p_-,r)$
to simplify the equation to
\eqn\bfbseq{
\mubf^2\tpbf(x)=\left({\alb1\over x}+{\alf2\over1-x}\right)\tpbf(x)
  -\pint{(x+y)\over2x}\tpbf(y)}
Here we defined $x\equiv p_-/r_-$,
$\alb i\equiv\pi\mb i^2/g^2\nc-1$, similarly for
$\alf i$ and $\mubf^2$ is the bound state mass squared
in units of $g^2\nc/\pi$.
${\rm P}\!\!\int dx$ denotes the principal value integral defined by
\eqn\ppint{{\rm P}\!\!\!\int\!\! dx\,
f(x)\equiv\half\int\!\!dx \biggl[f(x+i\epsilon)
+f(x-i\epsilon)\biggr]_{\epsilon\rightarrow0}}
For comparison, we also list the Bethe--Salpeter equations
for the fermion--fermion and the boson--boson bound states:
\eqn\bseq{\eqalign{
\muff^2\tpff(x) &=\left({\alf1\over x}+{\alf2\over1-x}\right)\tpff(x)
  -\pint\tpff(y)\cr
\mubb^2\tpbb(x) &=\left({\alb1\over x}+{\alb2\over1-x}\right)\tpbb(x)
  -\pint{(x+y)(2-x-y)\over 4x(1-x)}\tpbb(y)\cr
}}
It is important to note that while the expressions for the full propagators
in \props\ involve the infrared cutoff $\lm$, the bound state equations
\bfbseq,\bseq\ are independent of this cutoff.

The bound state equations in \bfbseq, \bseq\ are
essentially Schr\"odinger equations $\mu^2\varphi(x)=(H\varphi)(x)$.
The ``Hamiltonian'' operator in the case of the boson--fermion
(or the boson--boson) bound state equation on the right hand side is not
Hermitean with respect to the standard measure $\int_0^1dx$.
We may conjugate by $\sqrt x$ to define
\eqn\conjugate{\pbf(x)\equiv{\sqrt x}\tpbf(x)}
Then the bound state equation becomes Hermitean
with respect to the standard measure;
\eqn\hermeq{
\mubf^2\pbf(x)=\left({\alb1\over x}+{\alf2\over1-x}\right)\pbf(x)
  -\pint{(x+y)\over2\sqrt{xy}}\pbf(y)}
Application of the general arguments
given in \THOOFT\ref\VW{G.~Veneziano, \npb{\bf B117} (1976) 519;\nl
E. Witten, \npb{\bf B160} (1979) 57
and references therein.}
show that the widths of these ``meson'' bound states are of higher order,
of $\c O(\nf/\nc)$ and the spectrum is real.
The three point coupling between these  mesons is of
$\c O(1/\sqrt{\nc})$.

The behavior of the bound state wave function
at the boundaries may be obtained following \THOOFT.
If we denote the behavior of $\pbf$ as
\def\bb#1{\beta_{(b)#1}}
\def\bbf#1{\beta_{(f)#1}}
\eqn\bbbeq{\pbf(x)\buildrel x\sim0\over\sim x^{\bb1}
\qquad \pbf(x)\buildrel x\sim1\over\sim(1-x)^{\bbf2}}
then the boundary behavior at $x=0,1$ is determined by
the masses of the constituents from the equations
\eqn\bbeq{\alb1-\pi\bb1\tan(\pi\bb1)=0\qquad
\alf2+\pi\bbf2\cot(\pi\bbf2)=0}
The boundary behavior at $x=0$ and 1 is that of
the boson--boson and fermion--fermion bound state
wave functions respectively.
Physical considerations restrict the boundary conditions
to $\beta>0$ except for the case $m_f^2=0$,
when $\bbf{}=0$ is also allowed.

Approximate eigenstates may be obtained for the higher
mass  states as
\eqn\regge{{\varphi_{(bf)k}}\sim\sqrt2\sin(\pi k x)}
which has the bound state mass of approximately
$\pi^2|k|$ leading to a linear trajectory asymptotically.
This behavior is the same as that seen for the
fermion--fermion and the boson--boson bound states.

The boson--fermion bound state equation is a natural
generalization of the fermion--fermion, boson--boson
bound state equations derived previously.
In the large--$\nc$ limit in two dimensions,
the bound state is a ``meson'' formed out of
a scalar or a spinor quark and an anti--quark in a linear confining
potential.
{}From the point of view of applying the results of this work
to the standard model, the spectrum contains
no massless fermions even when the original theory
has chiral symmetry. This, in agreement with
the results derived in other approaches
\ref\SHROCK{I-H. Lee, R.~Shrock, \prl{\bf 59} (1987) 14;\nl
S.D.H.~Hsu, Harvard preprint,  HUTP--92--A047 (1993)},
is unlike what is seen in the real world.
However, to analyze the behavior
of the strongly coupled standard model,
we need to consider the effect of
the scalar quartic coupling as well.
We would also like to understand these bound states within
the string picture of two--dimensional QCD
\ref\STRINGQCD{
W.~Bardeen, I.~Bars, A.~Hanson, R.~Peccei, \prd{\bf D13} (1976) 2364;\nl
D.J.~Gross, Princeton preprint PUPT--1356 (1992);\nl
J.~Minahan, \prd{\bf D47} (1993) 3430;\nl
D.J.~Gross, W. Taylor, Princeton preprints PUPT--1376, PUPT--1382 (1993)}.
\listrefs
\end